# Precision measurements in nuclear $\beta$–decay with LPCTrap

G. Ban[1], D. Durand[1], X. Fléchard[1], E. Liénard[1] and O. Naviliat-Cuncic[2]

[1]LPC Caen, ENSICAEN, Université de Caen, CNRS/IN2P3, Caen, France
[2]NSCL and Department of Physics and Astronomy, Michigan State University, East-Lansing, MI, USA

Address: 6, Bd Maréchal Juin, 14050 Caen Cedex, France
Tel.: (33)231452420 Fax: (33)231452549
E-mail: lienard@lpccaen.in2p3.fr

## Abstract

The experimental achievements and the current program with the LPCTrap device installed at the LIRAT beam line of the SPIRAL1-GANIL facility are presented. The device is dedicated to the study of the weak interaction at low energy by means of precise measurements of the $\beta-\nu$ angular correlation parameter. Technical aspects as well as the main results are reviewed. The future program with new available beams is briefly discussed.

## 1. The $\beta-\nu$ angular correlation

The structure of the weak interaction remains an important topic in the area of fundamental interaction physics. In particular, at low energies, there are nowadays ambitious experimental programs to search for "exotic" currents beyond the *V-A* theory [1, 2]. Such studies have recently made significant progress with the advent of improved trapping techniques [3, 4]. These sophisticated setups allow the production of $\beta$ decay sources almost at rest and confined in small volumes which can be surrounded by suitable detectors [5, 6]. We describe here the achievements made with the LPCTrap setup installed at GANIL[1]. The experimental program is dedicated to precision measurements of the $\beta-\nu$ angular correlation parameter.

From a theoretical point of view, the generalization of Fermi's theory leads to consider five different Lorentz invariant contributions in the transition amplitude describing nuclear beta decay, which are referred to as *scalar* (*S*), *vector* (*V*), *tensor* (*T*), *axial-vector* (*A*) and *pseudoscalar* (*P*). The search for *S* and *T* exotic contributions can be performed through a precise measurement of the $\beta-\nu$ angular correlation parameter, $a_{\beta\nu}$. For allowed transitions and non-oriented nuclei, this parameter can be directly inferred from an events distribution linked to the angular correlation between the leptons [7]. Since neutrinos are difficult to detect, a sensitive observable for a $\beta-\nu$ angular correlation measurement is a kinematic parameter of the recoiling daughter nucleus. If the $\beta$ particles and the recoil ions are detected in coincidence and the relative time of flight (ToF) is measured between the two particles, the expected distribution of events in the electron energy and recoil ion time of flight, t, is for all solid angles detected:

$$N(T_e,t)dT_e dt = K\, F(\pm Z,E_e)(T_e+mc^2)(r(t))^3(Q-T_e)\xi\left(1+a\frac{(r(t))^2c^2-2T_eE_e-Q(Q-2T_e)}{2E_e(Q-T_e)}+b\frac{mc^2}{E_e}\right)dT_e dt \quad (1)$$

---

[1] GANIL: Grand Accélérateur National d'Ions Lourds

where $T_e$, $E_e$ and $r(t)$ denote the kinetic and total energies of the $\beta$ particle and the recoil ion momentum respectively. $Q$ is the energy available in the transition, $K$ is a constant and $F(\pm Z, E_e)$ is the Fermi function ($\beta^-$ and $\beta^+$ decays). The parameter $a_{\beta\nu}$ and the Fierz interference term $b$ depend on the coupling constants, $C_i$ and $C_i'$ ($i = S, V, T, A$), associated to the different contributions [7]:

$$a_{\beta\nu}\,\xi = \left|M_F\right|^2 [(-|C_S|^2 - |C_S'|^2 + |C_V|^2 + |C_V'|^2) \mp 2\frac{\alpha Zm}{p_e} \operatorname{Im}(C_S C_V^* + C_S' C_V'^*)]$$

$$+ \frac{|M_{GT}|^2}{3}[(|C_T|^2 + |C_T'|^2 - |C_A|^2 - |C_A'|^2) \pm 2\frac{\alpha Zm}{p_e} \operatorname{Im}(C_T C_A^* + C_T' C_A'^*)] \qquad (2)$$

$$b\,\xi = \pm 2\sqrt{1-\alpha^2 Z^2}\ \operatorname{Re}[|M_F|^2 (C_S C_V^* + C_S' C_V'^*) + |M_{GT}|^2(C_T C_A^* + C_T' C_A'^*)] \qquad (3)$$

where $\xi = |M_F|^2(|C_S|^2 + |C_S'|^2 + |C_V|^2 + |C_V'|^2) + |M_{GT}|^2(|C_T|^2 + |C_T'|^2 + |C_A|^2 + |C_A'|^2)$ \qquad (4)

$M_F$ and $M_{GT}$ are the Fermi (F) and Gamow-Teller (GT) nuclear matrix elements, $\alpha$ is the fine structure constant, $Z$ is the atomic number of the daughter nucleus and $m$ and $p_e$ are the mass and momentum of the $\beta$ particle respectively. The $P$ contribution is neglected in a non-relativistic description of the nucleons. The existence of both coupling constants, $C_i$ and $C_i'$, is related to the transformation properties under parity. The Standard Model (SM) assumes maximal parity violation ($|C_i| = |C_i'|$), time reversal invariance ($C_i$, $C_i'$ real) and pure vector and axial-vector interactions ($V$-$A$ theory). The $\beta$−$\nu$ angular correlation parameter is then given by:

$$a_{\beta\nu} = \frac{1-\rho^2/3}{1+\rho^2} \qquad (5)$$

where $\rho = \dfrac{C_A M_{GT}}{C_V M_F}$ is the mixing ratio of the transition,

leading to $a_{\beta\nu} = 1$ for a pure F transition and $a_{\beta\nu} = -1/3$ for a pure GT transition.

Under the conditions of the present experiment, the parameter which is determined experimentally is actually:

$$\tilde{a}_{\beta\nu} = a_{\beta\nu}/(1 + \left\langle b\frac{m}{E_e}\right\rangle) \qquad (6)$$

where the brackets $<>$ mean a weighted average over the measured part of the $\beta$ spectrum.

For mirror transitions, the measurement of $a_{\beta\nu}$ also allows for a precise determination of the mixing ratio $\rho$. If the half-life, the branching ratio ($BR$) and the masses involved in the transition are well known, this ratio can be used to deduce the first element of the Cabbibo-Kobayashi-Maskawa (CKM) quark mixing matrix, $V_{ud}$ [8]:

$$V_{ud}^2 = \frac{4.794\times 10^{-5}}{(Ft_{1/2})G_F^2|M_F|^2(1+\Delta_R)(1+\frac{f_A}{f_V}\rho^2)} \qquad (7)$$

where $Ft_{1/2}$ is the corrected ft-value of the transition, $G_F$ is the weak interaction fundamental constant, $\Delta_R$ is a common radiative correction and $f_A$ ($f_V$) is the rate statistical function computed for the GT (F) component. Many interesting mirror transitions exist [9], which could form an additional database to test the Conserved Vector Current (CVC) hypothesis in nuclei and to determine $V_{ud}$.

## 2.   The LPCTrap setup

The experimental setup is installed at the low-energy beam line LIRAT[2] of the SPIRAL1-GANIL[3] facility. Technical details of the setup are described elsewhere [5, 10-12]. The central element of the device is a transparent Paul trap (Fig. 1) constructed to confine singly charged radioactive ions, almost at rest, in a small volume, allowing the detection in coincidence of the $\beta$ particles and of the recoiling ions. The trap consists of 6 concentric rings with an open geometry, allowing easy injection and extraction as well as an efficient detection of the decay products.

Upstream of the trap, the installation is equipped with a Radio Frequency cooler-buncher (RFQCB) for the preparation of the beam by reducing the emittance and the production of ion bunches. The low energy radioactive beam is provided by the ECR source of the SPIRAL facility with a typical energy dispersion of about 20 eV at 10 keV total kinetic energy. The RFQCB is connected to the Paul trap by a short transport line with dedicated beam optics and diagnostics. The off line tuning of the ensemble is made possible by the coupling of an ion source at the entrance of the setup. There, a Faraday cup and a retractable silicon detector provide efficient beam diagnostics. A layout of the equipment is shown in Fig. 2.

The incident beam is first decelerated below a few tens eV by the H.V. applied to the RFQCB platform. This enables the cooling of the ions by collisions with a buffer gas ($H_2$ or $^4$He) at a pressure of the order of $10^{-2}$ mbar. Beam bunches are produced near the exit of the quadrupole at a repetition rate of the order of 10 Hz (cycle). A first pulse cavity is used to transport the bunches at 1 keV downstream from the RFQCB. A second pulse cavity reduces the kinetic energy of the ions at 100 eV for an efficient injection into the Paul trap. There, the ions are once again cooled down by elastic collisions with an inert gas injected at very low pressure ($10^{-6}$-$10^{-5}$ mbar). At equilibrium, the thermalized ions have energies of about 0.1 eV and the diameter of the ion cloud located at the centre of the trap is of the order of 2.5 mm (see Fig. 7 in next section).

Two different detection setups, shown in Fig. 3, have been used in the experiments. In both setups, a telescope and a micro-channel plate position sensitive detector (MCPPSD) detect the $\beta$ particles and the recoil ions respectively. In the first setup, both detectors are located 10cm away from the center of the trap in a back-to-back configuration. The $\beta$ telescope is made of a 60×60mm² 300μm thick double-sided silicon strip detector (DSSSD) for position readout, followed by a 7cm thick plastic scintillator coupled to a photomultiplier to measure the $\beta$ particle kinetic energy and to deliver the start signal for the recoil ion time of flight measurement. The stop signal is delivered by the MCPPSD whose performances are detailed in Ref. [13]. For each coincidence event, the positions of both particles, the recoil ion ToF, the $\beta$ particle kinetic energy, the time stamp of the decay within the cycle and the RF phase of the trap are recorded. This set of measured parameters is useful to control systematic effects and to check the consistency of the results.

In the second setup implemented in 2010 (Fig. 3, right panel), a spectrometer has been added to separate the charge states of the recoil ions. The ions emitted towards the spectrometer are accelerated by a -2kV potential after they cross the collimator located in front of a 50cm long free flight tube. An electrostatic lens set at -250V allows the collection of all the ions on the MCPPSD located at the end of the tube. The front plate of the detector is set at -4kV to ensure a maximum detection efficiency for all charge states, independently of the ion kinetic energy. This spectrometer makes LPCTrap a unique and unprecedented setup for the measurement of charge state distributions of ions associated with the $\beta$ decay of singly charged radioactive ions.

## 3. Measurements in the decay of $^6$He

The first nucleus of interest that has been considered is $^6$He, which decays via a pure GT transition. For such transitions, the experiment described here was the first performed after decades following the last measurements of the beta-neutrino angular correlation in the $\beta$ decay of $^6$He [14-16] and $^{23}$Ne [17, 18]. Another new experiment, measuring the $\beta-\alpha-\alpha$ correlation from trapped $^8$Li ions, has recently reported the first results [19]. Among these experiments, only one measurement in the decay of $^6$He [14] was performed with a relative precision at the level of 1%, yielding $a_{\beta\nu} = -0.3308(30)$ after inclusion of radiative and recoil-order currents corrections [20].

At GANIL, the $^6$He$^{1+}$ beam is produced by the fragmentation of a primary $^{13}$C beam at 75 MeV/u on a graphite target coupled with an ECR source. The low energy beam is delivered at 10 keV through the LIRAT beam line with a maximum intensity of about $1.5 \times 10^8$ pps. The resolving power (M/$\Delta$M ~ 250) of the dipole magnet located at the entrance of the low energy beam line is not sufficient to eliminate the 10 nA of a stable beam containing mainly $^{12}$C$^{2+}$, but this contamination is drastically reduced in the RFQ.

A first experiment was performed using the initial setup (Fig. 3, left panel) with a measurement cycle of 100ms. This experiment is described in details in [21]. The ToF spectrum of the recoil ions is fitted to extract $a_{\beta\nu}$. The analysis is based on the comparison between the experimental ToF spectrum and those obtained for two sets of realistic Monte Carlo (MC) simulations considering pure axial ($a_{\beta\nu} = -1/3$) and pure tensor ($a_{\beta\nu} = +1/3$) couplings. In a first step, the experimental data are calibrated and corrected for the identified sources of unwanted events (background). Then, $a_{\beta\nu}$ is deduced from a fit of the experimental ToF spectrum with a linear combination between the two sets of simulated decays obtained for axial and tensor couplings. Among the many instrumental effects, the detectors response function and geometry, the trap RF field influence, the ion cloud characteristics inside the trap, the shake-off ionization of the recoil ion, and the scattering of the $\beta$ particles are implemented in the simulations.

The calculation of the expected value of the angular correlation parameter in the SM includes radiative corrections [20, 22]. We have used the formalism described by Glück [20], based on the work of Sirlin [23] to calculate, to first order in $\alpha$ and on an event by event basis, the change in the kinematics due to the virtual and real photon emission during the decay process. It turns out that such corrections are at the 1% level on the value of the correlation parameter.

The shake off ionization of the recoil ion due to the sudden change of the electric charge of the nucleus following $\beta$ decay is also taken into account. For the present analysis, the shake off ionization probability of $^6$Li$^{2+}$ ions has been calculated in the sudden approximation, and found to be 0.02334 + 0.00004$E_{RI}$ where $E_{RI}$ is the ion recoil energy in keV. Since the maximum recoil energy is 1.4 keV, the energy dependent term can be neglected. This ionization probability is in perfect agreement with previous calculations of Wauters and Vaeck [24]. In a second experiment performed with the improved setup of fig.3, the shake-off probability was experimentally measured and was found to be in very good agreement with the theoretical value [12].

The characteristics of the ion cloud inside the trap have strong influence on the ToF spectrum. Simulations of the trapped ions trajectories in the Paul trap have thus been performed using the SIMION8 software package [25]. The geometry of the electrodes of the surrounding elements was included as well as the experimental RF trapping voltages. The collisions between the trapped ions and the $H_2$ buffer gas molecules were described at the microscopic level using realistic interaction potentials [26]. The position and velocity distributions at thermal equilibrium as a function of the RF phase can be obtained from such simulations. Based on these calculations, the mean thermal energy of the ion cloud is expected to be $kT_{sim} = 0.09$ eV. This is smaller than the value obtained in off-line measurements, $kT_{exp} = 0.107(7)$ eV [27]. A small correction (scaling) factor was thus applied to the cloud characteristics in the simulations. Measurements performed at different ion cloud densities have shown that space charge effects are negligible under the adopted measuring conditions. The response functions of the detectors dedicated to the detection of the electron and the recoil ion were simulated with particular emphasis on the influence of the backscattering of $\beta$ particles on the detectors and on other structures inside the detection chamber.

An important source that affects the shape of the ToF spectrum is the backscattering of electrons since, in this case, the kinematics of the decay is biased. The ToF distribution associated with such events ("scattered") is obtained by massive simulations using the GEANT4 software code.

False coincidences ("accidentals"), corresponding to the detection of uncorrelated particles on the $\beta$ telescope and on the micro-channel plate, also contribute to the ToF spectrum and introduce a flat distribution that is easily subtracted. It appears that most of such accidentals events originate from (i) the decaying neutralized $^6$He species which are present inside the whole volume of the chamber and trigger the $\beta$ telescope, and (ii) $H_2$ molecules leaking from the RFQCB which trigger the MCPPSD detector.

For a small fraction of the $^6$He atoms decaying in the chamber volume, the recoil ion can be detected on the MCPPSD, in coincidence with its associated $\beta$ particle. This constitutes another source of physical events labelled "out-trap" events. Their contribution to the ToF spectrum in the region of the fit can bias the measurement. They are included in the simulation, assuming a decay process with pure axial coupling.

The ToF spectrum obtained for validated coincidences is displayed in Fig. 4. The selected events are conditioned by: a 500 keV energy threshold on the energy deposited in the scintillator, a time within the trapping cycle between 25 and 95 ms (cloud at thermal equilibrium), and a valid reconstruction of the positions in both the DSSSD and the MCPPSD. The "scattered", "out-trap", and "accidentals" events represent respectively 4.5%, 2.6%, and 7.3% of the total number of events. These contributions and the ToF obtained for $a_{\beta\nu}$ = -1/3 are summed and compared to the experimental spectrum in Fig. 4.

By considering the $\beta$ decay vertex as a point like source at the centre of the Paul trap, the recoil ion ToF and position can be used to determine the three components of the recoil ion momentum. In a similar way, the full momentum vector of the $\beta$ particle can be deduced from the energy deposited in the $\beta$ telescope and the position on the DSSSD. This provides the possibility to reconstruct the antineutrino invariant mass: $m_\nu^2 = E_\nu^2 - p_\nu^2$

This variable is useful to validate in a global manner the quality of the simulations with a small influence of the value of the correlation parameter. Figure 5 shows the antineutrino invariant mass spectra obtained for the experimental and simulated events. The main peak is well reproduced by the simulations. The shape and position of this peak depend on all the inputs of the simulations (background, detector response functions, geometries, size of the ion cloud, trap RF field, etc…). The agreement between the experimental and simulated data provides an additional test of the inputs used in the analysis.

After applying a window cut in the antineutrino invariant mass spectrum and after background subtraction, the experimental ToF spectrum is adjusted with a linear combination of the time of flight spectra simulated using pure axial and pure tensor couplings (Fig. 6, left panel). Three parameters were left free in the fit: the value of $a_{\beta\nu}$, the total number of events, and the distance $d_{MCPPSD}$ between the MCPPSD detection plane and the centre of the Paul trap. Such a distance cannot be measured with better accuracy than a fraction of a mm and is therefore a degree of freedom to be taken into account in the analysis. The range in ToF selected for the fit is indicated by the vertical lines. The experimental data have first been split randomly in four independent sets, and the corresponding ToF spectra were fitted by changing the upper limit of the fitting range. No significant dependency has been found. Contours of constant $\chi^2$ in the plane of parameters $d_{MCPPSD}$ and $a_{\beta\nu}$ are shown on the right panel in Fig. 6. The result from the fit leads to $a_{\beta\nu} = -0.3335 \pm 0.0073$, and $d_{MCPPSD} = 100.255 \pm 0.011$ mm. The nominal value for this distance is 100.0 mm with a positioning uncertainty of 0.5 mm. The minimum chi-square $\chi^2_{min} = 96.6$ for 105 degrees of freedom corresponds to a P-value of 0.71 which indicates a very good agreement between the data and the fitted function. The quoted error is purely statistical.

The contributions of the main sources of systematic uncertainty are listed in Table 1. The label "present data" in the column "Method" of Table 1 indicates that the parameters and their uncertainties were determined by fitting the experimental data with the MC simulation (assuming a pure axial coupling). In each case, it was verified that the sensitivities of these parameters to the value of $a_{\beta\nu}$ taken as input were negligible at the current level of precision. The associated uncertainties on $a_{\beta\nu}$ were then deduced from the changes in the $a_{\beta\nu}$ values resulting from the fit of the experimental ToF spectrum while varying the parameters in the MC simulation.

It has been found that the $a_{\beta\nu}$ value resulting from the fit strongly depends on the trapped ion cloud size and on the temperature used in the MC simulation (Fig. 7). This motivated an independent measurement of the ion temperature using an off-line source of $^6Li^+$ [27], performed under identical running conditions than those in the $^6He^+$ experiment in terms of trap RF voltage, gas pressure in the trap chamber and number of trapped ions. A relative uncertainty of 6.5 % was obtained which constitutes the dominant contribution to the systematic error on the value of $a_{\beta\nu}$.

The uncertainties due to "accidentals" and "out-trap" event subtractions are only statistical. They are limited by the statistics of experimental events which serve as normalization for the MC results. The uncertainty due to the "scattered" events was estimated by considering a 10% relative error on the $\beta$ scattering yield provided by the GEANT4 simulations. This 10% relative error is based on the work of Hoedl [28], which compares a compilation of published electron scattering experimental data to several MC codes.

Combining all systematic errors quadratically, the final result is: $a_{\beta\nu} = -0.3335(73)(75)$ where the first uncertainty is statistical and the second systematic.

Figure 8 shows the first result obtained with LPCTrap compared with other measurements of the $\beta-\nu$ angular correlation coefficient in pure GT transitions. The values from Carlson and from Allen *et al.* were obtained in $^{23}Ne$ decay, the value of G. Li *et al.* was obtained in $^8Li$ decay and the others in $^6He$ decay. Our result and the one of G. Li *et al.* are the most accurate among the experiments performed using the detection of the recoil ions and $\beta$ particles in coincidence. The measurement presented here, performed with a different and independent method, confirms the result of Johnson *et al.* It is to be recalled that the reduced chi-square $\chi^2/\nu = 0.92$ for 105 degrees of freedom obtained in the present work corresponds to a P-value of 0.71. This is to be compared to a P-value of 0.055 for the Johnson *et al.* experiment, with a reduced chi-square $\chi^2/\nu = 1.69$ for 13 degrees of freedom [20].

The techniques used in the two experiments differ in a number of other aspects. First, the use of trapping techniques and the detection in coincidence of two decay products resulted in a larger signal to background ratio, by more than a factor of two compared to the Johnson *et al*. experiment. Furthermore, the measurement in an event by event mode of additional parameters (particle positions, energy of the $\beta$ particle, RF phase, and time within the trapping cycle) allows a better control of possible systematic effects. With the efficiencies achieved for beam preparation and trapping and for the detection of coincidence events, the average counting rates in the present experiment was about 1 coincidence per second and about 100 single triggers per second. This is respectively three and one orders of magnitude lower than in the Johnson *et al*. experiment. Thus, possible systematic effects due to the counting rate are expected here to have a smaller impact. The dominant contribution to the uncertainty in the Johnson *et al*. experiment was due to random variations of recoil energy spectra in a number of sequential data acquisitions, while in the present work, the precision limitation is mainly statistics. Both techniques are thus complementary.

From Eq.(6) and using the value $<m/E_e> = 0.2$ deduced from the $\beta$ particle energy spectrum [21], the result from the first experiment with LPCTrap corresponds to a limit on the tensor coupling

$|C_T/C_A| < 0.13$      (90% C.L.)

assuming $C_T = C_T'$.

A second experiment has been performed with the setup involving the recoil ion spectrometer (Fig. 3, right panel) [12]. A major improvement in the efficiency was reached resulting from several modifications of the setup. First, the injection inside the RFQ was optimized as well as the transmission between the RFQ and the trap. This led to the following efficiencies for a measurement cycle of 200 ms: 0.6% through the RFQ, 40% downstream up to the trap, 20% inside the trap. With an initial intensity of $10^8$ pps, this corresponds to about $10^4$ ions/bunch inside the Paul trap. The efficiency of detection was also increased by modifiying the geometry of the setup. All in all, a total of about $1.2 \times 10^6$ true coincidences were collected during four days of data taking.

In addition, the understanding of the systematic effect associated with the cloud temperature has been improved by studying in further details the behaviour of the leading edge of the ToF distribution [29]. The high statistics of the experiment also enables to improve the subtraction of background by carefully considering the different contributions of such events as a function of time inside the measurement cycle. In the second run, the ions were ejected from the trap after 150ms trapping duration to detect "out-trap" events during the 50 ms left.

As already stated above, an experimental determination of the shake-off probability was made possible with a very high accuracy. Figure 9 shows the experimental ToF distribution for the recoil ions detected in coincidence with the electron following the beta decay of $^6\text{He}^{1+}$ with the background subtraction described previously. In the present case, only two charge states are possible (Fig. 9, left). The fit, which takes into account all the effects mentioned above, is in perfect agreement with the experimental data. The value obtained for the shake-off probability is:

$P_{shake\text{-}off} = 0.02339(35)_{\text{stat}}(07)_{\text{syst}}$

The main sources of systematic effects have been investigated (Fig. 9, right), and the total systematic uncertainty is very small compared to the statistical error. The experimental value of $P_{shake\text{-}off}$ is in excellent agreement with the theoretical prediction based on the sudden approximation.

As far as the angular correlation parameter is concerned, a preliminary value has been estimated from another analysis: $a_{\beta v} = -0.3338(26)_{\text{stat}}$. The error on this value is dominated by the statistics used for the number of events in the simulation. The value is consistent with our previous result and the value obtained by Johnson *et al*.

The analysis of the experimental data is about to be finalized and the statistics gathered during the experiment should allow us to reach an unprecedented statistical accuracy of 0.0015.

## 4. Measurement in the decay of $^{35}$Ar

The $^{35}$Ar nucleus decays through a mirror transition with a large Fermi fraction $x = 92\%$. The basic parameters ($T_{1/2}$, $BR$, masses) involved in the decay being well known [9], a value of $a_{\beta\nu}$ with sufficient precision can be determined in the framework of the SM. The precise measurement of $a_{\beta\nu}$ in this transition can then be interpreted either to constrain scalar currents, or to determine $V_{ud}$. In a previous measurement, performed decades ago [17], the recoil ions after beta decay were detected in singles, leading to the value $a_{\beta\nu} = 0.97(14)$.

At the SPIRAL1-GANIL facility, the $^{35}$Ar$^{1+}$ ions are produced by a primary $^{36}$Ar beam at 80 MeV A$^{-1}$ impinging on a graphite target coupled to an ECR source. The radioactive beam is delivered to the LPCTrap at 10 keV with a typical intensity of $3.5\times10^7$ s$^{-1}$ (~ 5.5 pA), measured by the retractable silicon detector. The dipole magnet located at the entrance of the low energy beam line is again not sufficient to eliminate the 40 pA of a stable beam containing mainly molecules which have not been fully identified.

The tuning of the RFQ has been adapted to the $^{35}$Ar$^{1+}$ ions which are cooled down with He buffer gas at a $1.6\times10^{-2}$ mbar pressure. To prepare the experiments, tests were first carried out with different stable beams produced either by a surface ionization source or the ECR source of SPIRAL. Typical performances reached with the different beams are given in Table 2 for two values of the cycle (20 ms and 200 ms).

Due to space charge effects, the transmission, and consequently the lifetimes of the ions in the buncher and in the Paul trap, can be limited by the number of ions per bunch. For the values given in Table 2, the beam current at the entrance of LPCTrap was adjusted in each case to keep this number below $10^7$ ions per bunch in order not to saturate the RFQ. For a cycle of 200 ms, this number is reached with a typical beam intensity of 50 pA at the entrance of LPCTrap. A maximum of $2.5\times10^5$ ions can then be confined in the Paul trap. In the RFQ, the measured efficiencies are mainly defined by the losses at beam injection in the setup (the hole at the entrance has a 6 mm diameter) and by the losses due to charge exchange. A short cycle gives access to the first factor: roughly 70% of the beam is lost at 20ms. The values obtained for a longer cycle enable to determine the lifetime of the ions in the buncher due to trapping losses. This is of the order of 100 ms for the Ar$^+$ species and 250 ms for K$^+$ ions, which is consistent with a lower charge exchange expected with alkali ions. In the measurement Paul trap, the lifetime of confined ions has also been determined and is close to 500 ms for all species, with buffer He gas being injected in the chamber at a very low pressure.

During the last run performed with $^{35}$Ar$^{1+}$, an average of $2.5\times10^4$ radioactive ions were confined for each bunch injection, every 200 ms. Even if the total efficiency of LPCTrap is lower for the stable contaminant, the number of ions counted in the ion cloud monitor revealed that a total of $1.5\times10^5$ ions were actually confined in the Paul trap at each cycle. This does not jeopardize the experiment, but space charge effects should be considered in the data analysis. The trapped ions reached their thermal equilibrium about 20 ms after injection in the He buffer gas at a pressure of $1.5\times10^{-5}$ mbar. After 160 ms trapping duration dedicated to the correlation measurement, the ions were extracted towards the ion cloud monitor and "out-trap" events detected during the 40 ms left. The second detection setup involving the recoil ion spectrometer (Fig. 3, right panel) was used to measure the decay products and a total of $1.5\times10^6$ real coincidences were recorded. The ToF spectrum obtained during this run is shown in Fig. 10.

The different charge states of the $^{35}$Cl daughter ions are well separated by the recoil ion spectrometer. The charge state distribution had already been estimated from the commissioning run [30]. The preliminary values are: 1+: 75(1)%      2+: 17(0.5)%    3+: 6(0.5)%     4+: 1.5(0.5)%   5+: ~ 0.5%

The study of systematic effects is currently going on. Among these effects, the probability of charge-exchange collisions in the ambient gas, which was negligible in the case of $^6$Li in $H_2$ [12], has to be estimated in the present case involving many electrons.

The number of neutral recoil $^{35}$Cl atoms, which are not detected by the MCPPSD, can be deduced from the comparison between the number of expected coincidences (including neutrals) and the effective number of measured coincidences. The number of expected coincidences is simply computed using the number of $\beta$ particles detected in singles and the overall ion detection efficiency. This estimate leads to 72(10) % of neutral $^{35}$Cl recoils. Beyond the prototypical $^6$He$^{1+}$ case, this heavier system can reveal the role of more subtle shake-off dynamics involving many electrons as, for instance, the Auger processes [31].

The number of coincidences measured during this run corresponds to a statistical uncertainty of 0.002 on $a_{\beta\nu}$. In the study of the systematic effects, particular attention will also be devoted to the trap RF influence on the recoil motion. In the $^{35}$Ar decay, the recoil kinetic energy reaches at most 452 eV, which is about a factor of 3 lower than in the case of $^6$He. The RF effect could then become a dominant source of the systematic uncertainty. Moreover, as the total number of ions in the trap nearly reaches the trap capacity, the space charge effect has to be taken into account in the analysis. Finally, the second main branch of the decay with $\gamma$ emission ($BR = 1.23\%$) has also to be considered. Such systematic contributions should be properly managed and, assuming a systematic uncertainty of the same order than the statistical one, the final result would constitute the most precise value ever obtained in a $\beta-\nu$ angular correlation measurement. On one hand, this would add a relevant contribution to better constrain the scalar interaction. On the other hand, in the framework of the SM, this result would improve by a factor of 2 the precision on $V_{ud}$ determined from the study of mirror transitions.

## 5. Summary and perspectives

The status of the first precision experiments in nuclear $\beta$ decay performed with LPCTrap at GANIL has been described. Such studies are based on the use of a Paul trap to confine the radioactive ions and on the detection in coincidence of the $\beta$ particles and the recoil ions. Significant results, especially the most precise value of $a_{\beta\nu}$ ever determined for a pure GT transition using a coincidence method, have been obtained. The last data are very promising and they should deliver the most precise absolute results on both the tensor and scalar sectors. A significant improvement of $V_{ud}$ extracted from the mirror transitions also seems reachable in an alternative interpretation of the $^{35}$Ar data. Moreover, an original recoil ion spectrometer enabled to measure, for the first time, the charge state distributions of recoiling ions produced by the decay of 1+ ions. The use of the sudden approximation in the theoretical approach has definitively been validated with the result obtained with the 'ideal' $^6$He$^{1+}$ system, allowing unambiguous tests of the role of other processes in systems involving many electrons as in $^{35}$Ar$^{1+}$ decay.

A new project involving $^{19}$Ne is now underway. This nucleus decays mainly through a mirror transition to the ground state of $^{19}$F ($BR = 99.9858(20)\%$). The basic parameters ($T_{1/2}$, $BR$, masses) involved in the decay being well known [9, 32], a value of $a_{\beta\nu}$ can be determined in the framework of the SM with sufficient precision. The precise measurement of $a_{\beta\nu}$ in this transition can again be interpreted either to constrain the exotic currents, $C_S$ and $C_T$, in a global way, or to determine $V_{ud}$. This experiment is a new challenge for LPCTrap. The half-life of $^{19}$Ne is indeed ten times larger than in the case of $^{35}$Ar and the recoil maximum kinetic energy is only 203 eV. These are extreme conditions that will push the limits of the device to new levels of sensitivity.

Very exciting opportunities will emerge in the coming years at GANIL with the development of new radioactive beams in an upgrade of SPIRAL [33] and in the forthcoming DESIR[4] hall at SPIRAL2 [34]. The intensities reached during the first production tests for nuclei such as $^{23}$Mg, $^{25}$Al, $^{33}$Cl and $^{37}$K, which decay through mirror transitions, are encouraging and these beams should be available at GANIL in 2015.

---

[4] DESIR: Décroissance Excitation et Stockage d'Ions Radioactifs

Tables

| Source | Uncertainty | $\Delta a_{\beta\nu}$ (x $10^{-3}$) | Method |
|---|---|---|---|
| Cloud temperature | 6.5% | 6.8 | off-line measurement |
| "accidentals" and "out trap" | see text | 0.9 | present data |
| $\beta$ Scattering | 10% | 1.9 | GEANT4 |
| Total (see [21]) | | 7.5 | |

Table 1:   Main sources of systematic error, systematic uncertainties, and impact on the error of $a_{\beta\nu}$ that are discussed in the text. The total uncertainty quoted corresponds to the sum of all effects taken into account in the analysis for which some of them are not discussed here.

| Species | Source | ε(RFQ) | | ε(transfer line) | ε(trap) | ε(total) | |
|---|---|---|---|---|---|---|---|
| | | 20 ms | 200 ms | | | 20 ms | 200 ms |
| $^{39}K^{1+}$ | Off-line | 0.27 | 0.17 | 0.43 | 0.2 | 0.023 | 0.015 |
| $^{36,40}Ar^{1+}$ | ECR/SPIRAL | 0.32 | 0.12 | 0.22 | 0.15 | 0.01 | 0.0037 |
| $^{35}Ar^{1+}$ | ECR/SPIRAL | / | 0.15 | 0.25 | 0.10 | / | 0.0038 |

Table 2:   Efficiencies of the three main sections of LPCTrap obtained with different stable and radioactive ions around mass 35, for two values of the cycle (20ms and 200ms).

Figures

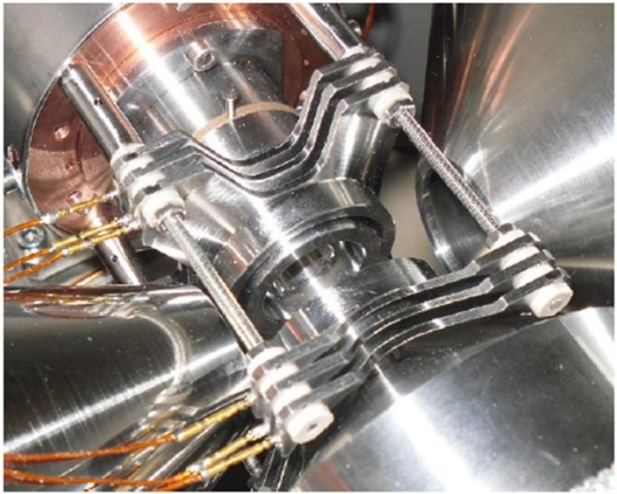

Figure 1: Picture of the transparent Paul trap of the LPCTrap setup

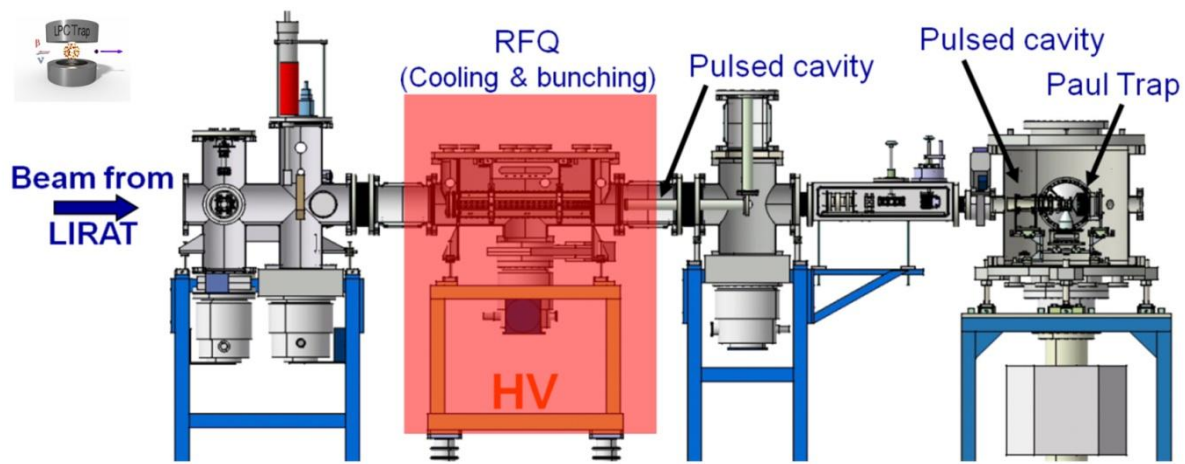

Figure 2: Scheme of the full LPCTrap setup

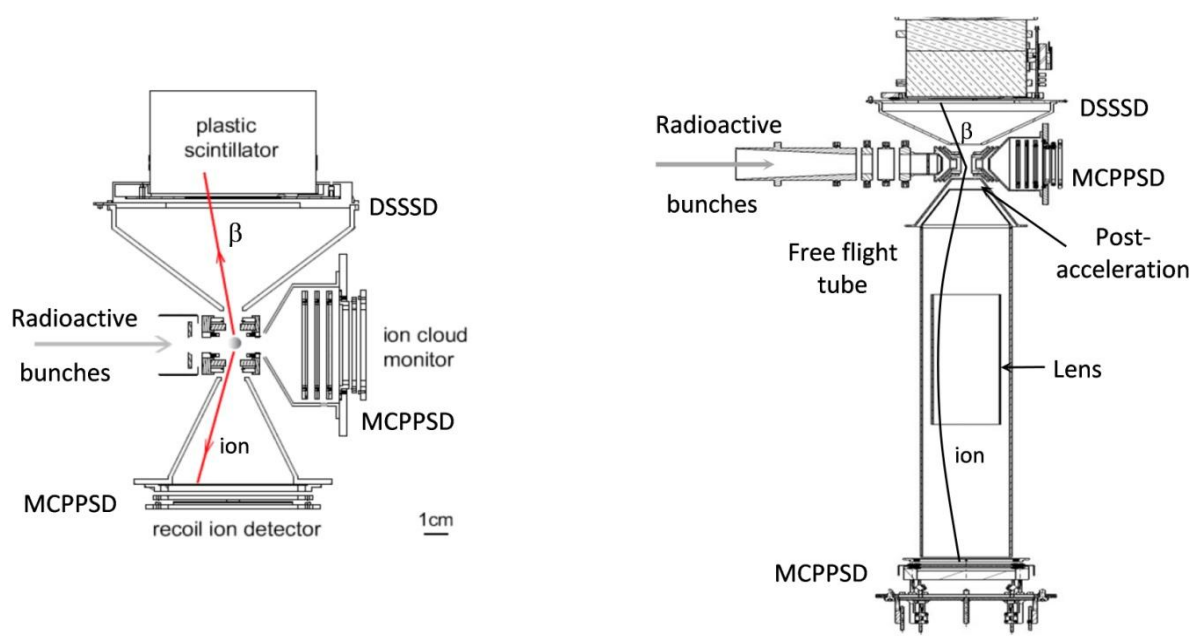

Figure 3: Top view of the detection chamber of LPCTrap. *Left panel*: First detection setup used in the commissioning phase. *Right panel*: Improved detection setup allowing the separation of the charge states of the recoil ions.

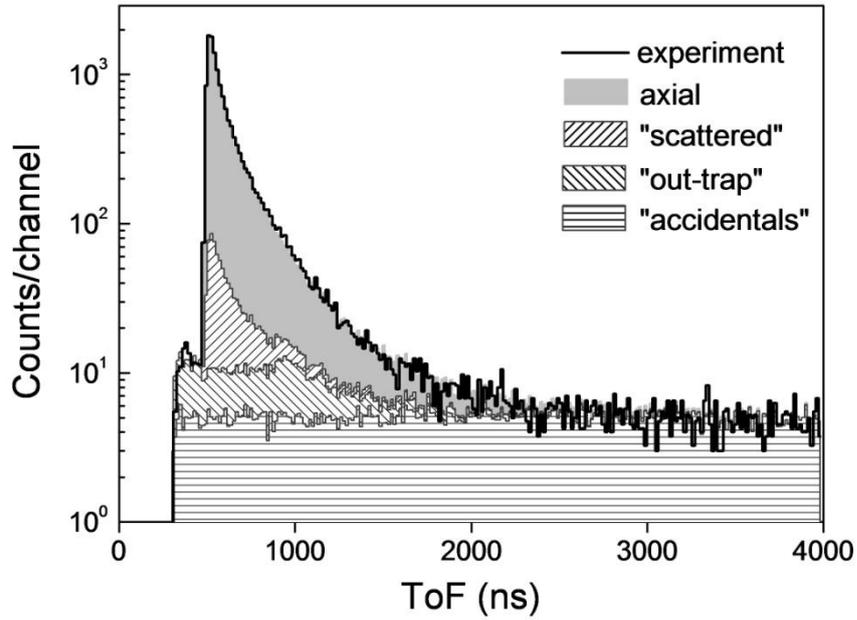

Figure 4: Experimental time of flight spectrum (black line) compared to the simulated one (grey area) in the pure axial case, including the different simulated contributions of background, after normalization (see text for details).

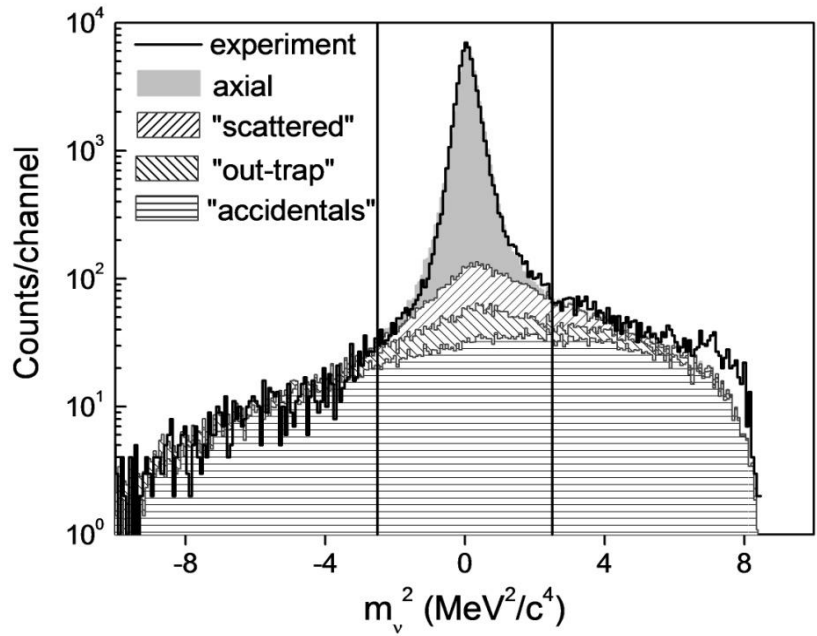

Figure 5: Antineutrino invariant mass spectra for the experimental data, the simulation in the pure axial case, and the simulation of background events. Vertical lines indicate the selection window used in the analysis to reduce background contribution.

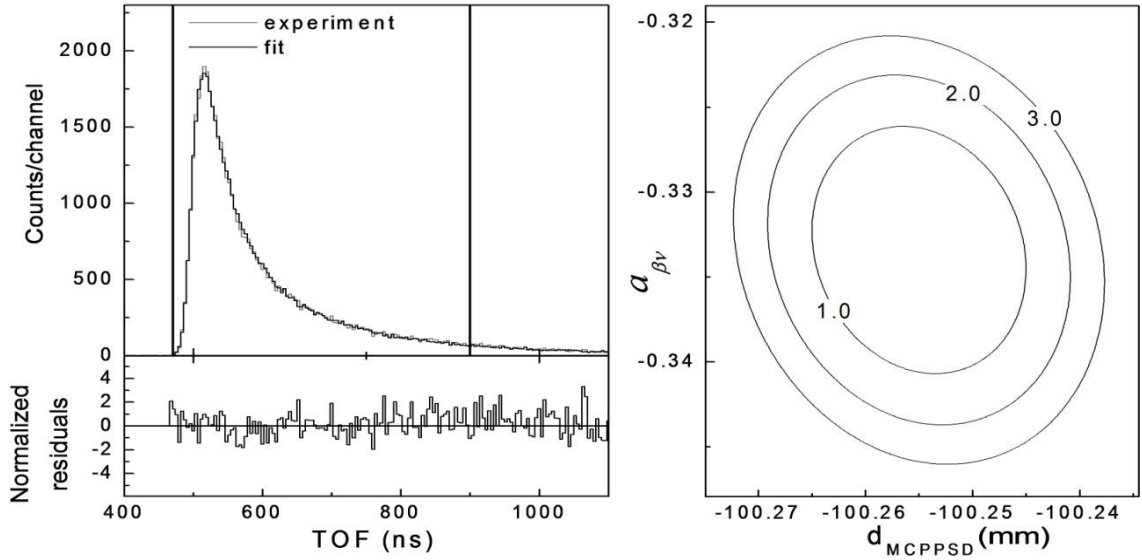

Figure 6: *Left panel:* Fit of the experimental spectrum (upper panel). The range of the fitted function is indicated by vertical lines. The normalized residuals are plotted in the lower panel. *Right panel:* projection on the plane of parameters $d_{MCPPSD}$ and $a_{\beta\nu}$ of the computed contours for ($\chi^2 - \chi^2_{min}$) values = 1, 2, and 3.

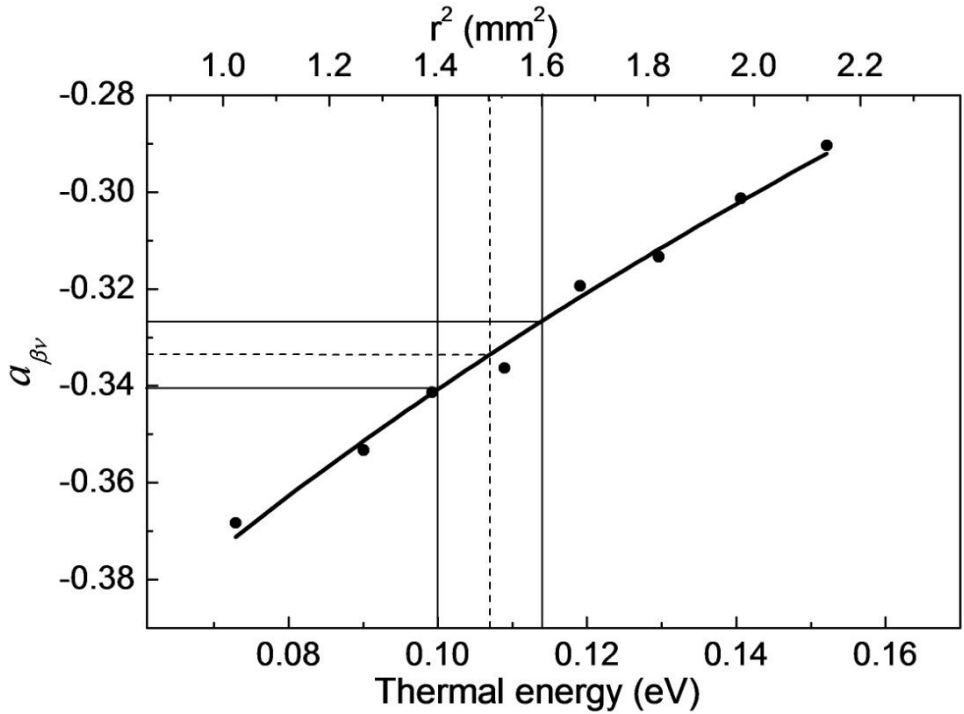

Figure 7: Values of the correlation coefficient resulting from the fit procedure as a function of the ion cloud thermal energy, $kT$, and of the ion cloud square radius, $r^2$, where $r$ is the RMS of the spatial distribution in the Paul trap radial plane. The dashed and solid lines correspond respectively to the central value and the $1\sigma$ uncertainty of the off-line temperature measurement.

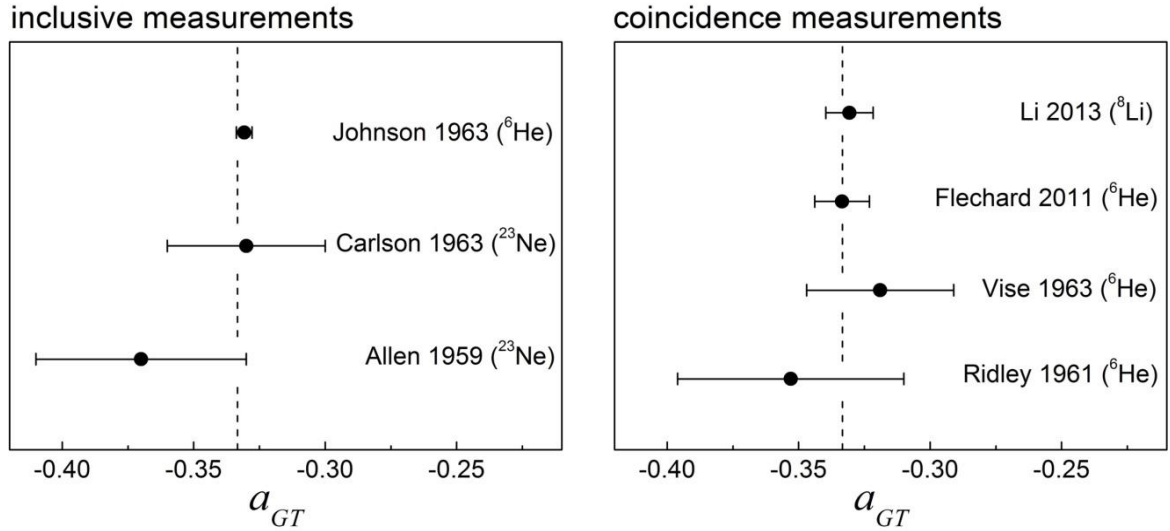

Figure 8: From top to bottom: $a_{\beta\nu}$ experimental values in pure GT transitions from [14], [18], [17] (left panel), [19], present work, [15], and [16] (right panel). The error bars show the quadratic sums of statistical and systematic uncertainties. The dashed lines indicate the value predicted by the SM.

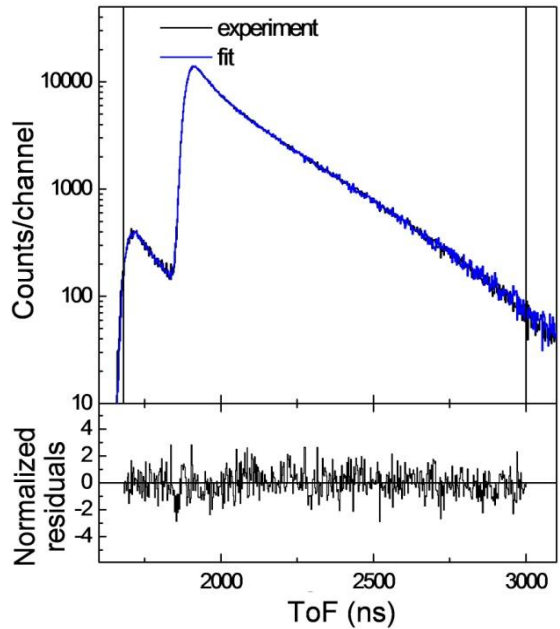

| Source | $\Delta P(10^{-5})$ | Method |
|---|---|---|
| $a_{\beta\nu}$ | 4.0 | [19] |
| $\beta$ scattering | 4.0 | GEANT4 |
| Background | 3.5 | present data |
| $E_\beta$ calibration | 1.7 | present data |
| MCP efficiency | 1.2 | present data |
| Total | 7.0 | |

Figure 9: *Left panel*: Fit of the experimental spectrum to deduce the charge state distribution. *Right panel*: Dominant source of systematic effects.

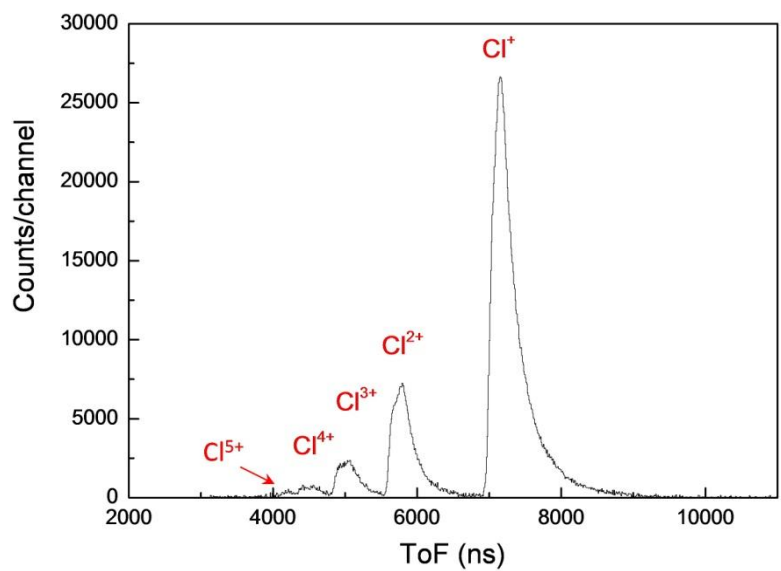

Figure 10: Experimental ToF spectrum obtained during the run with $^{35}$Ar, for about $1.5 \times 10^{6}$ coincidence events.